\documentclass[intlimits,twoside,a4paper]{article}
\usepackage{amsmath,amssymb}
\usepackage{graphicx}

\usepackage[T2A]{fontenc}
\usepackage[cp1251]{inputenc}

\usepackage{graphics}

\newcommand{\Dt}{\frac{\mbox{d}}{\mbox{d}t}}

\usepackage[eqsecnum]{cmpj3}

\articletype{Rapid communication}

\issue{2017}{20}{4}{44001}
\doinumber{10.5488/CMP.20.44001}
\title[Synchronization with anchoring in Smorodinsky-Winternitz potentials]%
{Synchronization and anchoring of two non-harmonic canonical-dissipative 
oscillators via Smorodinsky-Winternitz potentials}
\author[S. Mongkolsakulvong, T.D. Frank]{S. Mongkolsakulvong\refaddr{label1}, 
T.D. Frank\refaddr{label2,label3}}
\addresses{
\addr{label1}
Faculty of Science, Department of Physics, Kasetsart University, Bangkok
10900, Thailand\\
\addr{label2} 
CESPA, Department of Psychology, 
University of 
Connecticut, 406 Babbidge Road, Storrs, CT 06269, USA\\
\addr{label3} 
Department of Physics,
University of 
Connecticut, 2152 Hillside Road, Storrs, CT 06269, USA
}

\date{Received July 18, 2017}
\begin{document}

\maketitle

\begin{abstract}
Two non-harmonic canonical-dissipative limit cycle oscillators are considered
that oscillate in one-dimensional Smorodinsky-Winternitz potentials. It is
shown that the standard approach of the canonical-dissipative framework to
introduce dissipative forces leads naturally to a coupling force between the
oscillators that establishes synchronization. The non-harmonic character of
the limit cycles in the context of anchoring, the phase difference between the
synchronized oscillators, and the degree
of synchronization are studied in detail.

\keywords canonical-dissipative systems, 
Smorodinsky-Winternitz potentials,
synchronization, anchoring

\pacs 
 05.45.-a, 
05.40.Jc, 
87.19.rs 
\end{abstract}

Mass-spring systems are a fundamental topic of
classical mechanics and solid state physics. A sophisticated theoretical
framework is available to explain the oscillatory dynamics of 
 macroscopic particles connected
by springs and the oscillatory vibration of molecules 
interacting by spring-like forces. Given the success
of this field of physics concerned with the inanimate world,
the question naturally arises whether its scope can be
broadened to take the life science into account~\cite{pikovsky01book}.
Since a key feature of living systems is their ability to move by themselves, 
the question can be asked from a slightly different perspective. Can the
concepts of classical mechanics be generalized to self-mobile, so-called 
active~\cite{schweitzer03book,lindner08epjst,frank14cmp}, systems? 
A well-studied class of
active systems both in the animate and inanimate world 
are self-oscillators~\cite{jenkins13physrep}.
A theoretical framework that bridges the research fields of 
classical mechanics and 
self-oscillators is the theory of 
canonical-dissipative (CD) 
systems~\cite{haken73zeitphys,graham73a,ebeling04cmp,ebeling04book,mongkolsakulvong10cmp,romanczuk12epjst}.
The reason for this is that a CD system can exhibit attractors that, 
on the one hand, are stable and in doing so reflect non-conservative, 
dissipative system components
but, on the other hand, are defined in their respective phase
spaces by the dynamics of conservative systems.   
In fact, in a series of recent experimental studies it has been shown that
the 
CD approach can be applied to human self-oscillators, that is, humans
producing oscillatory single limb 
movements~\cite{dotov11mc,dotov15biosystems,kimseokhun15}. Importantly, this
line of research has been generalized to the non-harmonic 
case~\cite{gordon16cmp}. In general, human rhythmic limb movements exhibit 
non-harmonic components and in particular can show a so-called anchoring
phenomenon. Anchoring means that a limb movement slows down 
during a particular short period of the cyclic activity
in a 
more pronounced way than in the harmonic case. Non-harmonic self-oscillator
models are promising candidates to capture human non-harmonic rhythmic
activity including the anchoring phenomenon.
However, humans and animals are known to coordinate
their activities, in general, and movement patterns, in 
particular~\cite{schmidt90jephpp,winfree01book}. As far as the CD approach is
concerned, for synchronization with zero phase lag and 180~degrees phase lag,
a four-variable CD model has been proposed
recently~\cite{chaikhan16actamech}. 
Here, we proposed a more general model for two
active particles that is
motivated by the so-called SET model for swarming~\cite{schweitzer01pre}
and assumes that active particles are
coupled via their angular momentum values 
(see also~\cite{ebeling04physd}).  In order
to address the non-harmonic case, we consider self-oscillations in
Smorodinsky-Winternitz 
potentials~\cite{friswinternitz65,winternitz67,tegmen04ijmpa}. 
These potentials
play an important role in physics as confinement 
potentials~\cite{jin13,jasmine15}. 
In the context of the CD
approach, the four-dimensional, two particle 
Smorodinsky-Winternitz systems should be considered as benchmark
systems because they feature three invariants rather than only two. 
The first two invariants are
the particle Hamiltonian energy functions. The third invariant is an
appropriately adjusted angular momentum~\cite{tegmen04ijmpa}.

Recall that the
standard CD 
oscillator in a one-dimensional space
with coordinate $q$ and momentum $p$ 
is defined 
by~\cite{ebeling04cmp,ebeling04book,frank05book,romanczuk12epjst}
\begin{equation}
\label{eq1}
\Dt q=\frac{p}{m}\,,  \qquad \Dt p=-k q -\frac{\partial g}{\partial p} \, , \qquad 
g=\frac{\gamma}{2} (H-B)^2 , 
\end{equation}
where $m$ denotes mass, $k$ is the spring constant and
$H=p^2/(2m)+k q^2/2$ corresponds to the Hamiltonian energy in the conservative
case in which the function $g$ is neglected. 
The function $g$ describes the dissipative mechanism. 
The parameter $\gamma\geqslant 0$ is the coupling parameter of
oscillator with the
dissipative mechanism. Note that $-\partial g/\partial p=-\gamma p(H-B)/m$
such that
the dissipative mechanism is composed of a negative
friction term (i.e., pumping mechanism) $+\gamma B p/m$ 
with pumping parameter $B\geqslant  0$ and a nonlinear
friction term $-\gamma p H/m$. The amplitude dynamics can be obtained using
standard techniques. 
We put
$q=A(t)\exp(\ri\omega t) +\text{c.c.}$ with $\omega^2=k/m$, where $A$ denotes the
complex-valued oscillator amplitude that is related to the real-valued
amplitude $r(t)$ and the oscillator phase $\phi(t)$ like
$A(t)=r(t)\exp[\ri\phi(t)]/2$. Here and in what follows, c.c. denotes the
complex conjugate expression.
Assuming that $\gamma$ is a small perturbation parameter,  
by means of the slowly varying amplitude
approximation and the rotating wave approximation~\cite{haken85book}, we obtain,
in lowest order of $\gamma$, the following amplitude dynamics
\begin{equation}
\label{eq2}
\Dt A= -\gamma \omega^2 A \left(|A|^2-\frac{B}{2m\omega^2}\right) . 
\end{equation}
Note that higher order correction terms in $\gamma$ can be obtained using 
alternative techniques (see e.g., \cite{omalley06}). From equation~(\ref{eq2})
it follows that
$\mbox{d} r/\mbox{d}t=- \gamma \omega^2 
r [ r^2-2B/(m \omega^2)]/4$
and $\mbox{d}\phi/\mbox{d}t=0$.
The solution reads
$r(t)=\sqrt{
2B r_0^2/[m\omega^2 r_0^2+(2B-m\omega^2 r_0^2)\exp(-\gamma B t/m)]
}$ with $r_0=r(t=0)$. Figure~\ref{fig1}~(a) 
shows a simulation of the self-oscillator~(\ref{eq2})
and the analytical solution $r(t)$. From equations~(\ref{eq2}) and the
analytical solution $r(t)$ it
 follows that $\gamma$ (in combination with the factor $B/m$) 
determines the time scale of the amplitude dynamics
$A(t)$ and $r(t)$, respectively. Importantly, in the long time limit, $r$
approaches $r_{\rm st}=\sqrt{2B/(m\omega^2)}$ and $H$ converges to the pumping
parameter $B$. That is, $B$ acts as a fixed point value or target value for the
energy dynamics $H(t)$.

\begin{figure}[htb]
\centerline{\includegraphics[width=0.8\textwidth]{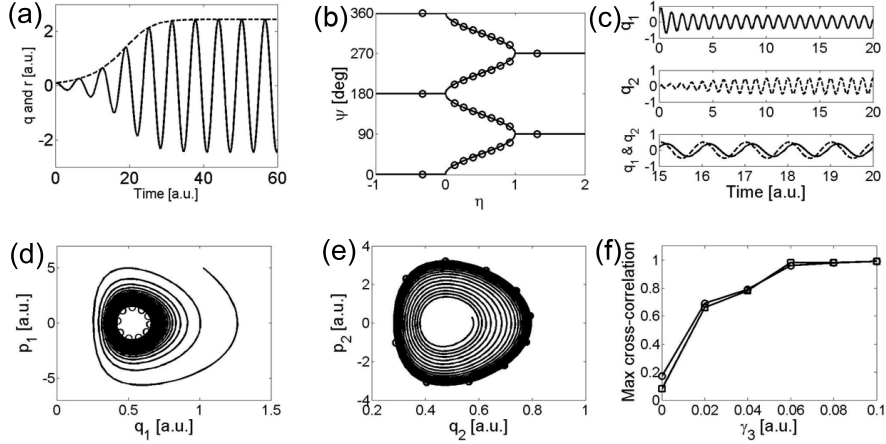}}
\caption{Panel~(a): 
Solution $q(t)$ (solid line) of equation~(\ref{eq1}) 
obtained by a numerical Euler forward (EF) solution method. 
The analytical solution
  $r(t)$ is shown as well (dashed line). Parameters in a.u.: $m=k=1$, 
$\gamma=0.1$, $B=3$. EF time step: $\tau=0.01$.
Initial value: $q(0)=0.1$, $p(0)=0 \Rightarrow
  r(0)=0.1$. 
Panel~(b):
Phase difference
  $\psi$ as function of $\eta$ 
  (solid lines) as predicted by our theoretical considerations 
for the harmonic case (see text). 
Circles denote simulation results obtained by solving equation~(\ref{eq6})
numerically~(EF). Simulation parameters in a.u.: 
$m=1$, $k=(2\piup)^2 \Rightarrow \omega=2\piup$ (i.e., oscillator period equal to 1
time unit), $\alpha=\beta=0$, $B_1=3$, $B_2=5$, 
$B_3$ was varied in the range
  $[-0.5,1.5]$, $\tau=0.001$. 
 Various initial conditions were used. 
Panel~(c): Trajectories $q_1(t)$ and $q_2(t)$ of 
equation~(\ref{eq6}) for a representative simulation trial used to generate the
numerical results in panel~(b). Top and middle sub-panels show transient and long
term dynamics. Bottom sub-panel shows the synchronized state with a fixed
phase difference. Here: $B_1=1.0 \Rightarrow \eta=0.66 \ , \ \psi=54^\circ$.
Panels~(d) and (e): Phase portraits $p_1$ versus $q_1$ [panel~(d)] and  
$p_2$ versus $q_2$ [panel~(e)] obtained in the non-harmonic case by solving 
equation~(\ref{eq6}) numerically (EF). 
Parameters in a.u.:  
$m=1$, $k=(2\piup)^2 \Rightarrow \omega=4\piup$ (i.e. 
oscillator period equal to 0.5 time units), 
$\alpha=3$, $\beta=2$, $\gamma_{1,2}=0.1$, $\gamma_3=0.2$, 
$B_1=H_{1,\rm min}+1$, $B_2=H_{2,\rm min}+5$, $B_3=S_{3,\text{min}}+3$,
$\tau=0.001$.
Initial
conditions: $p_1=5.0$, $p_2=0.3$, $q_1=(\alpha/k)^{0.25}+0.5$, 
$q_2=(\beta/k)^{0.25}+0.1$. 
The circles show the predicted
limit cycles obtained by solving numerically (EF) 
the evolution equations of 
the corresponding isolated, conservative
oscillators. Panel~(f): Maximal cross-correlation as function of
$\gamma_3$ obtained by solving equation~(\ref{eq16}) numerically (stochastic 
EF~\cite{risken89book}). 
Averages of 10 trials are shown. Trajectories of 10000
(circles) and 30000 (squares) time units were used in each trial. The maximal
cross correlation scores for $\gamma_3=0$ are by-chance values that decay to
zero when trajectory length goes to infinity. Parameters in a.u.: $D=0.02$, all
other parameters except for $\gamma_3$ as in
panels~(d) and (e). $\tau=0.001$.
} \label{fig1}
\end{figure}

Let us generalize the single-oscillator case to a model of 
two coupled self-oscillators. Each self-oscillator oscillates in a
one-dimensional space and is 
subjected to the force of a 
Smorodinsky-Winternitz potential. Let us describe the oscillator coordinates
$q_1$ and $q_2$ and momenta $p_1$ and $p_2$ by means of the vectors 
${\bf q}=(q_1,q_2)$ and
${\bf p}=(p_1,p_2)$.
The 
Smorodinsky-Winternitz potentials read~\cite{tegmen04ijmpa}
\begin{equation}
V_1(q_1)=k\frac{q_1^2}{2}+\frac{\alpha}{2 q_1^2} \, , \qquad
V_2(q_2)=k\frac{q_2^2}{2}+\frac{\beta}{2 q_2^2} 
\end{equation}
with $k>0$ and 
$\alpha,\beta \geqslant  0$. For $\alpha=\beta=0$, the potentials correspond to
parabolic potentials and $k$ can be interpreted as a spring constant. 
For $\alpha,\beta>0$, the potentials exhibit minima at 
$q_1=\pm(\alpha/k)^{1/4}$ 
and 
$q_2=\pm(\beta/k)^{1/4}$ and exhibit repulsive singularities 
at $q_1=0$ and $q_2=0$. By
contrast, for $q_1\to \pm \infty$ and $q_2\to \pm \infty$ 
they increase like
parabolic potentials. Therefore, for $\alpha,\beta>0$, 
the potentials are asymmetric with respect to
their minima.
Let us define the dynamics of the two oscillators in the conservative case
by means of the Hamiltonian dynamics
\begin{equation}
\label{eq4}
\Dt {\bf q} = \frac{\partial H_{\rm tot}}{\partial {\bf p}} \, , \quad
\Dt {\bf p} = -\frac{\partial H_{\rm tot}}{\partial {\bf q}} \, , \quad
H_{\rm tot}=H_1+H_2 \, , \quad
H_1= \frac{p_1^2}{2m} + V_1 \, , \quad
H_2= \frac{p_2^2}{2m} + V_2 \, .
\end{equation}
In equation~(\ref{eq4}), the functions $H_1$, $H_2$ and $H_{\rm tot}$
correspond to the Hamiltonian energy functions of the individual oscillators 
and
the total energy of the two-oscillators system.
It can be shown that the dynamics~(\ref{eq4}) exhibits
three invariants $S_j$ with $j=1,2,3$ given by~\cite{tegmen04ijmpa}
\begin{equation}
S_1=H_1 \, , \qquad S_2=H_2 \, , \qquad
S_3=\frac{L^2}{m}+(q_1^2+q_2^2)
\left(\frac{\alpha}{q_1^2}+\frac{\beta}{q_2^2}\right) , 
\end{equation}
where $L$ denotes the
angular momentum $L=p_2 q_1 - p_1 q_2$. In line 
with the CD oscillator~(\ref{eq1}), we
define the 
CD case like
\begin{equation}
\label{eq6}
\Dt {\bf q} = \frac{\partial H}{\partial {\bf p}} \, , \qquad
\Dt {\bf p} = -\frac{\partial H}{\partial {\bf q}} - \frac{\partial
  g_{\rm tot}}{\partial {\bf p}} \, , \qquad
g_{\rm tot}=\sum_{j=1}^3 g_j \, , \qquad g_j = 
\gamma_j \frac{1}{2}\left(S_j - B_j\right)^2, 
\end{equation}
where $\gamma_j\geqslant  0$ are the coupling constants. 
$B_{1,2}\geqslant  0$ are the pumping
parameters. $B_3$ is a target value for~$S_3$. 
Importantly, 
since $S_j$ are invariants of the conservative dynamics~(\ref{eq4}), 
for the dissipative dynamics~(\ref{eq6}) it
follows that $\mbox{d}g_{\rm tot}/\mbox{d}t 
= -(\partial g_{\rm tot}/\partial {\bf p})^2 \geqslant  0$.
In view of the boundedness of $g_{\rm tot}$ (i.e., $g_{\rm tot}\geqslant  0$), 
$g_{\rm tot}$ 
is a Lyapunov function that becomes stationary in the long term limit. This in
turn implies that $\partial g_{\rm tot}/\partial {\bf p}=0$ 
such that equation~(\ref{eq6})
reduces to equation~(\ref{eq4}). In total, for $t\to \infty$, the
system~(\ref{eq6}) converges to an attractor that
corresponds to a solution of the conservative system~(\ref{eq4}).

Let us consider the harmonic case defined by $\alpha=\beta=0$. 
Using
$q_k=A_k(t)\exp(\ri\omega t) +\text{c.c.}$ with $\omega^2=k/m$ again and
$A_k(t)=r_k(t)\exp[\ri\phi_k(t)]/2$ and assuming that $\gamma_j$ are small
perturbation parameters, we obtain
\begin{eqnarray}
&& \Dt A_1= 
-\gamma_1 \omega^2 A_1 \left(|A_1|^2-\frac{B_1}{2m\omega^2}\right) 
- \ri \frac{\gamma_3}{m^2\omega} A_2 \,  U(A_1,A_2) , \nonumber\\
&& \Dt A_2= 
- \gamma_2 \omega^2 A_2 \left(|A_2|^2-\frac{B_2}{2m\omega^2}\right) 
+ \ri \frac{\gamma_3}{m^2\omega} A_1 \, U(A_1,A_2) 
\end{eqnarray}
with
$
U(A_1,A_2)=L (L^2/m-B_3)$ 
and 
$ L=2m \ri\omega 
(A_1^* A_2-A_1 A_2^*)$.
In terms of the real-valued amplitudes $r_1$ and $r_2$ 
and the phase difference 
$\psi=\phi_1-\phi_2$, we obtain
\begin{eqnarray}
\label{eq10}
\Dt r_1 =- \frac{\gamma_1 \omega^2}{4} r_1 
\left( r^2_1-\frac{2B_1}{m\omega^2} \right)
-\frac{\gamma_3 r_2}{m^2 \omega} \sin(\psi) \, U(\psi,r_1,r_2),\\
\Dt r_2 =- \frac{\gamma_2 \omega^2}{4} r_2 
\left( r^2_2-\frac{2B_2}{m\omega^2} \right)
-\frac{\gamma_3 r_1}{m^2 \omega^2} \sin(\psi) \, U(\psi,r_1,r_2)
\end{eqnarray}
with
$U(\psi,r_1,r_2) = m \omega r_1 r_2 \sin(\psi)
(L^2/m-B_3)$,  $L=m \omega r_1 r_2 \sin(\psi)$, and
\begin{equation}
\label{eq12}
\Dt \psi = -\frac{\gamma_3}{m^2 \omega} \left(
\frac{r_1^2+r_2^2}{r_1 r_2}\right) \cos(\psi) \, U(\psi,r_1,r_2) .
\end{equation}
A detailed stability 
analysis based on equations~(\ref{eq10})--(\ref{eq12}) 
for the case that both oscillators are excited like $B_1,B_2>0$, 
shows that the target level $B_3$ defines the location of the aforementioned
attractor. By rescaling $B_3$ we obtain 
the location parameter $\eta=m\omega^2 B_3/(4B_1 B_2)$.
It can be shown that for 
$B_3 \leqslant 0 \Rightarrow \eta \leqslant 0$, the two-oscillators system exhibits a stable
attractor characterized by $S_3=0 \Rightarrow \psi=0^\circ 
\ \vee \ \psi=180^\circ$
and $H_{1,2}=B_{1,2}$. This implies that
 $g_{\rm tot} \to \gamma_3 B_3^2/2>0$ for $t\to
\infty$. 
For $B_3\geqslant  0$, we distinguish between two cases. 
If $\eta \in [0,1]$, then
the limit cycle attractor is characterized by 
$H_{1,2}=B_{1,2}$ again but with $S_3=B_3$. The latter relation
implies that the phase difference is given by 
$\psi=\arcsin(\sqrt{\eta})$,
$\psi=180^\circ-\arcsin(\sqrt{\eta})$,
$\psi=\arcsin(\sqrt{\eta})+180^\circ$,
and
$\psi=360^\circ-\arcsin(\sqrt{\eta})$.
Moreover, 
 $g_{\rm tot} \to 0$ for $t\to
\infty$. 
For $B_3>0$ and $\eta>1$, it is impossible to have 
$H_{1,2}=B_{1,2}$ and $S_3=B_3$. Rather, the attractor is given by 
$\psi=90^\circ \ \vee \ \psi=270^\circ$, $H_{1,2}=B_{1,2}$ and $S_3<B_3$.
In addition, we have 
 $g_{\rm tot} \to \gamma_3 (S_3-B_3)^2/2 = 8\gamma_3 
B_1^2B_2^2(1-\eta)^2/(m^2\omega^4)>0$ 
for $t\to
\infty$. 
Figure~\ref{fig1}~(b) illustrates the attractor location in terms of the 
phase difference 
$\psi$ as a function of the location parameter $\eta$. Figure~\ref{fig1}~(c) 
shows, for a
representative simulation, the trajectories $q_1$ and $q_2$. The trajectories 
demonstrate that the two-oscillators system converges to a stable periodic
pattern (limit cycle).

Let us consider the 
non-harmonic case with $\alpha,\beta>0$. First we note that  the individual
oscillators (i.e., $\gamma_{1,2,3}=0$) for small amplitudes experience a
linearized force of $f(q_j)=-k_{\rm lin} q_j$ with $k_{\rm lin}=4 k$
irrespective of $\alpha$ and $\beta$. Consequently, the oscillation frequency
is two times the oscillation frequency of the harmonic case and the period is
half the period of the harmonic case. This implies that when removing the
singularities in the potentials $V_j$ by putting $\alpha=\beta=0$, then the
oscillation frequency drops in a discontinuous fashion from $2\omega_0$ to
$\omega_0$ with $\omega_0^2=k/m$. 
In general, the oscillation 
period for the oscillators $j=1,2$ can be
computed from the integral $T_j=2\sqrt{m}
\int_{ q_{\rm min} }^{ q_{\rm max} } \{2[ H_j(t=0)-V_j(q_j)]\}^{-1/2}
\mbox{d}q_j$ with $H_j(t=0)$ being the initial energy of oscillator $j$.
The integration limits
$q_{\rm min,max}$ are the turning points defined by $V_j=H_j(t=0)$. 
Numerical computations show that
$T_j$ is independent of $H_j(t=0)$ and 
$\alpha$ and~$\beta$. 
In line with the small amplitude oscillation case, we obtain
$T_j=T_0/2 \ \forall \ H_j(t=0)\geqslant  H_{j,\rm min}$, $\alpha,\beta>0$ with
$T_0=2\piup/\omega_0$, where $H_{j,\rm min}$ denote 
the minimal
energy values  $H_{1,\rm min}=\sqrt{\alpha k}$ and  
$H_{2,\rm min}=\sqrt{\beta k}$. 

Let us consider the case $\gamma_{1,2,3}>0$.
For the sake of brevity, we consider only the case in which all three invariants
of the conservative dynamics,
$H_{1,2}$ and $S_3$,
converge to their respective target values (i.e., $H_{1,2} \to B_{1,2}$
and $S_3\to B_3$) such that
$g_{\rm tot} \to 0$.
Our first objective is to
 show that the shapes of the oscillator limit cycles are distorted compared to
the harmonic case. Panels~(d) and (e) of figure~\ref{fig1} show 
the phase portraits $q_1,p_1$ and
$q_2,p_2$ obtained from a numerical simulation. The trajectories (solid lines)
converge to
``egg shaped'' limit cycles. The limit cycles are defined by the limit
cycles of the corresponding conservative oscillators with $\gamma_{1,2,3}=0$. 
For the example shown in panels~(d) and (e), the limit cycles of the corresponding 
conservative oscillators are illustrated by circles. 
Importantly, the limit cycles reveal an anchoring phenomenon. 
The dynamics slows down (more pronounced as in the 
harmonic
case) when the oscillators swing to the right of 
their
potential
minimum locations $q_1=(\alpha/k)^{0.25}$  and $q_2=(\beta/k)^{0.25}$, see
figures~\ref{fig1}~(d) and (e). 
By contrast, the dynamics speeds
up when the oscillators swing to the left of their potential minimum
locations. 
This is because the forces are
relatively weak on the right-hand sides [where $V_j(q_j)$ are
approximatively 
parabolic potentials] and relatively strong on the
left-hand sides [where $V_j(q_j)$ exhibit singularities].

Our second objective is to address the synchronization of the oscillators.
On the limit cycles, the oscillators oscillate with the same oscillation 
frequency of $\omega=2\omega_0$, see above.
From
$H_{1,2} \to B_{1,2}$ and $S_3\to B_3$ 
it follows that
\begin{equation}
\label{eq13}
\frac{p_1^2}{2m} = B_1 - V_1(q_1)  , \qquad
\frac{p_2^2}{2m} = B_2 - V_2(q_2)  , \qquad
B_3=\underbrace{\frac{L^2}{m}+(q_1^2+q_2^2)
\left(\frac{\alpha}{q_1^2}+\frac{\beta}{q_2^2}\right)}_{\displaystyle W}.
\end{equation}
Using two first equations of (\ref{eq13}), the function $L$ occurring 
in $W$ (defined above) 
can be expressed as
\begin{equation}
L(q_1,q_2)=p_2 q_1 - p_1 q_2 =
(-1)^n q_1 \sqrt{2m[B_2-V_2(q_2)]}
+(-1)^m q_2 \sqrt{2m[B_1-V_1(q_1)]}
\end{equation}
with
$m,n \in \{0,1\}$. This implies that the last equation of (\ref{eq13}) can be written as
$B_3=W(q_1,q_2)$.
The 
synchronized state is then described by
\begin{equation}
\Big\{
m \ddot{q}_1 = - \frac{\mbox{d}}{\mbox{d}q_1} V_1(q_1) \ \Rightarrow \ q_1(t) 
\Big\}
\ \wedge \
\Big\{ 
B_3=W(q_1,q_2) \ \Rightarrow \ q_2(t) 
\Big\} . 
\end{equation}
That is, the coordinate $q_2$ is given by a nonlinear (implicit)
mapping from $q_1$ to $q_2$. Importantly, this mapping is stable 
against perturbation because the two-oscillators system is attracted to 
the state with $g_{\rm tot}=0$ 
and $S_{1,2,3}=B_{1,2,3}$. Therefore, the two oscillators
are synchronized. Note that the same argument holds in the opposite
direction. Considering the second oscillator as independent oscillator,
the coordinate $q_1$ of the first oscillator is given by a nonlinear
(implicit) mapping from $q_2$ to $q_1$.

Let us illustrate the synchronization of the two oscillators by 
considering the CD oscillator model~(\ref{eq6}) under the impact of
fluctuating forces. Using a standard approach for introducing noise terms 
into 
CD systems~\cite{haken73zeitphys,graham73a,ebeling04book,romanczuk12epjst},
equation~(\ref{eq6}) becomes
\begin{equation}
\label{eq16}
\Dt {\bf q} = \frac{\partial H}{\partial {\bf p}} \, , \qquad
\Dt {\bf p} = -\frac{\partial H}{\partial {\bf q}} - \frac{\partial
  g}{\partial {\bf p}}  + \sqrt{D}
\left(
\begin{array}{l}
\Gamma_1(t)\\
\Gamma_2(t)
\end{array} 
\right) , 
\end{equation}
where $\Gamma_j(t)$ are independent Langevin forces~\cite{risken89book} 
normalized to 2 units. 
The parameter $D\geqslant  0$
is the diffusion constant. The Langevin equation~(\ref{eq16}) exhibits a
Fokker-Planck equation that can be cast into the form of a 
free energy Fokker-Planck equation~\cite{frank05book}. The stationary
probability density $P(q_1,q_2,p_1,p_2)$ can then be expressed in terms of 
a Boltzmann function of $g_{\rm tot}$ 
as $P=\exp(- g_{\rm tot}/D)/Z_0$, where $Z_0$ is a
normalization factor~\cite{schweitzer01pre,frank05book}. Considering 
$\gamma_j$ as small perturbation parameters, we may introduce the smallness
parameter $\gamma_0$ and put $\gamma_j =c_j \gamma_0$ with $c_j\geqslant 0$. 
Then, $P=\exp(-{\tilde g}_{\rm tot}/\theta)/Z_0$ 
with ${\tilde g}_{\rm tot}=g_{\rm tot}/\gamma_0$ holds, 
where
$\theta=D/\gamma_0$ can be considered as a non-equilibrium temperature.
In this form, the analogy to equilibrium systems becomes 
obvious~\cite{haken73zeitphys}.
Importantly, without coupling between the oscillators, that is, for 
$\gamma_3=0 \Rightarrow c_3=0$, we have $P(q_1,q_2,p_1,p_2)=
P_1(q_1,p_1)P_2(q_2,p_2)$ with $P_j=\exp(-g_j/D)/Z_j$ for $j=1,2$, 
where $Z_j$ are
normalization factors again. That is, the probability density
$P(q_1,q_2,p_1,p_2)$
 factorizes.
By analogy, the transition probability density (conditional probability
density) factorizes. Therefore, for $\gamma_3=0$, 
there are no cross-correlations between the
oscillators at any time lag.

Let us show with the help of
a stochastic CD model~(\ref{eq16}) 
that for $\gamma_3>0$, the two-oscillators model exhibits
a
stable synchronized state. 
To this end, we 
solved numerically equation~(\ref{eq16}) for a fixed value of $\gamma_3$ and
calculated the cross-correlation coefficients 
$\mbox{Corr}\big(q_1(t),q_2(t-\tau)\big)$ 
for different time lags $\tau \in [0,T=T_0/2]$.
We determined the maximal coefficient. Subsequently, we varied $\gamma_3$.
In doing so, we obtain the maximal cross-correlation coefficient as a  
function of the coupling parameter
$\gamma_3$. Figure~\ref{fig1}~(f) summarizes the simulation results. For
$\gamma_3=0$, there was a finite by-chance 
value for the maximal cross-correlation
coefficient that decayed when the simulation duration was increased.
As far as the impact of $\gamma_3$ is concerned, figure~\ref{fig1}~(f)
demonstrates that 
the maximal cross-correlation increased as a function of $\gamma_3$ --- as
predicted. This increase of the maximal cross-correlation coefficient 
was taken as the evidence
that for $\gamma_3$, the two oscillators were to some degree synchronized.

Future studies may focus, in particular, on the stochastic aspects of the
proposed CD two-oscillators model. For example, it has been suggested to use the analytical
solution for the short-time propagator to define maximum likelihood estimators
that can be used to estimate the model 
parameters of CD systems~\cite{frank10physletta}. In
fact, for single CD oscillator models in a series of studies, the CD 
theory has been applied to the experiments on human rhythmic motor behavior 
and model parameters have been estimated from experimental data both in the 
harmonic~\cite{dotov11mc,dotov15biosystems,kimseokhun15}
and non-harmonic case~\cite{gordon16cmp}.

\ukrainianpart

\title{Синхронізація та анкерування  двох  негармонічних канонічно-дисипативних
осциляторів  за допомогою потенціалів  Смородинського-Вінтерніца}
\author{С. Монгколсакувонг\refaddr{label1}, Т.Д. Франк\refaddr{label2,label3}}
\addresses{
\addr{label1}
Факультет природничих наук, відділення фізики, університет Касертсарт, Бангкок
10900, Таїланд\\
\addr{label2}
CESPA, відділення психології,  Коннектикутський університет, CT 06269, США\\
\addr{label3}
Відділення фізики, Коннектикутський університет, CT 06269, США
}
\makeukrtitle

\begin{abstract}
Розглянуто канонічно-дисипативні граничні цикли для двох негармонічних
осциляторів, що коливаються в одномірних потенціалах
Смородинського-Вінтерніца. Показано, що стандартний канонічно-дисипативний
підхід із введенням дисипативних сил природно приводить до появи взаємодії між
осциляторами, яка синхронізує їх рух. Детально досліджено негармонічний
характер граничних циклів у контексті анкерування, різницю фаз між
синхронізованими осциляторами та ступінь їх синхронізації.

\keywords канонічно-дисипативні системи, потенціали
Смородинського-Вінтерніца, синхронізація, анкерування

\end{abstract} 
\end{document}